\documentclass[aps,prd,superscriptaddress,twocolumn,nofootinbib,10pt]{revtex4}
\usepackage[pdftex]{color}
\usepackage[dvipsnames]{xcolor}
\usepackage[sort&compress]{natbib}
\usepackage[colorlinks=true,linkcolor=OrangeRed,filecolor=blue,urlcolor=Plum,citecolor=Plum,pdftex,plainpages=false]{hyperref}
\usepackage{url}
\pdfoutput=1
\usepackage{ifpdf}
\usepackage{color,hhline,psfrag,rotating}
\usepackage{dcolumn}
\usepackage{verbatim}
\usepackage{bm}
\usepackage{slashed}
\usepackage{graphicx}
\usepackage{amsfonts,amssymb,amsmath}
\usepackage{booktabs}  % Improves tables - needed for iPython LaTeX output.

\begin{document}
\title{Renormalisation of the tensor current in lattice QCD and the $J/\psi$ tensor decay constant}

\author{D.~Hatton}
\email[]{d.hatton.1@research.gla.ac.uk}
\affiliation{SUPA, School of Physics and Astronomy, University of Glasgow, Glasgow, G12 8QQ, UK}
\author{C.~T.~H.~Davies}
\email[]{christine.davies@glasgow.ac.uk}
\affiliation{SUPA, School of Physics and Astronomy, University of Glasgow, Glasgow, G12 8QQ, UK}
\author{G.~P.~Lepage}
\affiliation{Laboratory for Elementary-Particle Physics, Cornell University, Ithaca, New York 14853, USA}
\author{A.~T.~Lytle}
\affiliation{INFN, Sezione di Roma Tor Vergata, Via della Ricerca Scientifica 1, 00133 Roma RM, Italy}
\collaboration{HPQCD collaboration}
\homepage{http://www.physics.gla.ac.uk/HPQCD}
\noaffiliation

\date{\today}

\begin{abstract}
Lattice QCD calculations of form factors for rare Standard Model processes such as $B \to K \ell^+ \ell^-$ 
use tensor currents that require renormalisation. These renormalisation factors, $Z_T$, have
typically been calculated within perturbation theory and the estimated uncertainties from missing 
higher order terms are significant. Here we study tensor current renormalisation using 
lattice implementations of momentum-subtraction schemes. Such schemes are potentially more accurate 
but have systematic errors from nonperturbative artefacts. 
To determine and remove these condensate contributions we calculate the 
ground-state charmonium tensor decay constant, $f_{J/\psi}^T$, which is also of interest in 
beyond the Standard Model studies. We obtain 
$f_{J/\psi}^T(\overline{\text{MS}}, 2\ \mathrm{GeV})=0.3927(27)$ 
GeV, with ratio to the vector decay constant of 0.9569(52), significantly below 1. 
We also give $Z_T$ factors, converted to the $\overline{\mathrm{MS}}$ scheme, corrected 
for condensate contamination. 
This contamination reaches 1.5\% at a renormalisation scale of 2 GeV (in the preferred RI-SMOM 
scheme) and so must be removed for 
accurate results.
\end{abstract}

\maketitle
\section{Introduction}
\label{sec:intro}

Rare Standard Model processes, for example those that first appear at 1-loop order through
so-called ``penguin" diagrams, are of great interest in searches for new physics. 
The very low rate for the process in the Standard Model means that beyond the Standard Model searches have small 
backgrounds. The signal rate will also be small, however, so it is important to have firm theoretical understanding 
of the Standard Model contribution. 
This starts with the effective weak Hamiltonian, $\mathcal{H}_{\mathrm{eff}}$, after integrating out the weak bosons. 
$\mathcal{H}_{\mathrm{eff}}$ contains flavour-changing neutral- current 
operators that can induce, for example, rare $b \rightarrow s$ processes~\cite{Hurth:2010tk}.
Sandwiched between hadronic states these operators yield matrix elements 
that can be converted into form factors for differential decay rates for 
comparison to experiment. 
The best way to calculate the matrix elements is by using lattice QCD. 
The matrix elements required are those of operators in a continuum 
scheme for QCD, however, ideally in the same scheme in which the Wilson coefficients for 
the operators in $\mathcal{H}_{\mathrm{eff}}$ were determined (the 
$\overline{\mathrm{MS}}$ scheme). 
This means that the lattice operators must be matched accurately 
to the continuum scheme. For such $b\rightarrow s$ processes 
tensor operators in $\mathcal{H}_{\mathrm{eff}}$, e.g. 
$\overline{s}_L \sigma^{\mu\nu} b_R$, cause a particular problem 
for lattice to continuum renormalisation, because they cannot be connected 
to conserved currents. We show how to solve that problem here. 

An example of a rare $b \rightarrow s$ process being studied experimentally 
is $B \to K \ell^+ \ell^-$ decay. A first unquenched
lattice QCD calculation of this decay was performed in \cite{Bouchard:2013pna} by members of
the HPQCD collaboration and another in \cite{Bailey:2015dka} by the Fermilab lattice
and MILC collaborations. The former used Highly Improved Staggered Quark (HISQ) \cite{Follana:2006rc} light and strange quarks and NRQCD $b$ quarks and
the latter used asqtad light and strange quarks and Fermilab $b$ quarks. In the HPQCD calculation
the tensor current was renormalised using one-loop lattice QCD perturbation theory 
for the NRQCD-HISQ current. A 4\% systematic
uncertainty on the tensor form factor was then taken to account for missing 
higher order terms in $\alpha_s$. The
Fermilab/MILC calculation also used one-loop lattice QCD perturbation 
theory for the Fermilab clover-asqtad 
current renormalisation. The $\mathcal{O}(\alpha_s^2)$ uncertainty on the tensor form factor was taken 
as 2\%. 

The HPQCD collaboration has recently performed a series of $b$ physics calculations using the
HISQ formalism for all quarks, working upwards in mass from that of the $c$ quark 
and mapping out the dependence on the heavy-quark mass
\cite{McNeile:2011ng,McNeile:2012qf,McLean:2019sds,McLean:2019qcx}. The
success of this methodology indicates the possibility of improvement on previous $B \to K$
calculations for which it would be important also to reduce the uncertainty arising from the
tensor current renormalisation.

Here we use a partially nonperturbative procedure for the renormalisation 
using momentum-subtraction schemes implemented on the lattice as an intermediate
scheme~\cite{Martinelli:1994ty}. This produces tensor current renormalisation factors
with better accuracy than those used in the calculations mentioned above because the perturbative part
of the calculation, the matching from momentum-subtraction to the $\overline{\mathrm{MS}}$ scheme, 
can be done through $\alpha_s^3$ in the continuum.
Renormalisation factors calculated on the lattice in momentum-subtraction schemes 
suffer from nonperturbative artefacts in general. 
Because these survive the continuum limit they need to be removed or otherwise accounted for. 
The artefacts 
are suppressed by powers of the renormalisation scale $\mu$ and can therefore be studied by 
performing calculations at multiple $\mu$ values, as we did for the quark mass 
renormalisation in~\cite{Lytle:2018evc}.
We show here how to remove such systematic effects in the tensor renormalisation factor by 
calculating a simple matrix element of the tensor operator 
that we can determine accurately in the continuum limit.
For this purpose we use the $J/\psi$ tensor decay constant $f_{J/\psi}^T$.

The vector $J/\psi$ decay constant $f_{J/\psi}^V$, calculated from the vector charmonium correlator, 
is related to the leptonic decay rate of the $J/\psi$ meson. For a recent 
very accurate determination of this decay constant see~\cite{Hatton:2020qhk}.
In contrast there is no simple decay rate that can be related to the $J/\psi$ tensor
decay constant. 2-flavour lattice QCD and QCD sum rules calculations of $f_{J/\psi}^T$ and the
ratio $f_{J/\psi}^T/f_{J/\psi}^V$ were presented in~\cite{Becirevic:2013bsa}, and we
will compare to those results here. 
$f_{J/\psi}^T$ is required for the calculation of bounds on beyond the Standard
Model charged lepton flavour violating $J/\psi$ decay rates~\cite{Hazard:2016fnc} and a similar 
calculation for other vector mesons would extend this. The $B_s^*$
tensor decay constant appears in parameterisations of its Standard Model decay rates
$B_s^* \to \ell^+ \ell^-$~\cite{Grinstein:2015aua}. Calculation of this decay constant 
is underway using the tensor current renormalisation factors we have determined here. 

In the next Section we discuss the definition of the tensor current renormalisation 
factor in the RI-SMOM and RI$'$-MOM momentum-subtraction schemes. In Section~\ref{sec:ZT-latt} 
we give details of our lattice calculation of the tensor renormalisation factor. 
This is followed by our lattice calculation of the $J/\psi$ tensor decay 
constant in Section~\ref{sec:jpsi-tensor}. Our results 
for $f_{J/\psi}^T$ are discussed 
in Section~\ref{sec:discussion-f} followed by discussion of our $Z_T$ results in Section~\ref{sec:discussion}. 
Finally, we give our conclusions in Section~\ref{sec:conclusions}.

\section{$Z_T$ in the RI-SMOM and RI$'$-MOM schemes} \label{sec:ZT-intro}

Momentum-subtraction schemes provide useful intermediate schemes in matching 
lattice QCD to the continuum $\overline{\text{MS}}$ scheme because they provide a way
to implement the same scheme both on the lattice and in the continuum~\cite{Martinelli:1994ty}. 
Then the continuum limit of the lattice results will be in 
the continuum momentum-subtraction scheme (and independent of the lattice action used) 
and can be matched to the $\overline{\text{MS}}$ 
in continuum QCD. 
 
In both of the momentum-subtraction schemes that we consider here the wavefunction renormalisation 
$Z_q$ is defined in terms of the inverse of the momentum space quark propagator $S(p)$ 
according to \cite{Martinelli:1994ty,Chetyrkin:1999pq,Aoki:2007xm,Sturm:2009kb}
\begin{equation} \label{eq:Zq-def}
    Z_q = - \frac{1}{12p^2} \mathrm{Tr}[S^{-1}(p)\slashed{p}] .
\end{equation}
As the propagator is gauge-dependent it is necessary to work in a fixed gauge. Landau 
gauge is used throughout.
Working in a fixed gauge raises the possibility of effects from
Gribov copies. Here we do not address this issue and assume that such effects are negligible
following general expectations and the findings of \cite{Gattringer:2004iv}, which saw no observable effects at a
precision below 1\%.

The tensor current renormalisation is defined in terms of $Z_q$ and the tensor vertex
function $G_T$:
\begin{equation} \label{eq:GT}
\begin{split}
  &G_T(p_1,p_2) = \\ &\int d^4x d^4y_1 d^4y_2 e^{iqx} e^{-ip_1 y_1} e^{ip_2 y_2} \langle T^{\mu\nu}(x) \overline{\psi}(y_1) \psi(y_2) \rangle .
\end{split}
\end{equation}
Here $T^{\mu\nu}(x)$ is the tensor current $\overline{\psi}(x) \sigma^{\mu\nu} \psi(x)$. 
We take the bilinears in the renormalisation procedure to
be non-diagonal in flavour. The renormalisation of flavour singlet and non-singlet
tensor bilinears are the same on the lattice through at least two-loop level and we may therefore safely use the $Z_T$ calculated here for
any flavour structure of the tensor current \cite{Constantinou:2016ieh}.

\begin{table}
\caption{Matching factors and tensor current running factors required to match our lattice results 
to the $\overline{\mathrm{MS}}$ scheme at a scale of 2 GeV. The second column gives the conversion factor between the RI-SMOM and
$\overline{\mathrm{MS}}$ schemes for the various $\mu$ values in the first column. 
The RI$'$-MOM to $\overline{\mathrm{MS}}$ matching factors are given in the third column. 
The factor that accounts for $\overline{\mathrm{MS}}$ running to a scale of
2 GeV for different values of $\mu$ is given in the fourth column. This used the 
three-loop tensor anomalous dimension
from~\cite{Gracey:2000am}. All of these values are correlated through
their use of a common determination of $\alpha_s$, taken from~\cite{Chakraborty:2014aca}.}
\label{tab:convs}
\centering
\begin{tabular}{llll}
\hline \hline
$\mu$ [GeV] & $Z_T^{\overline{\mathrm{MS}}/\mathrm{SMOM}}(\mu)$ & $Z_T^{\overline{\mathrm{MS}}/\mathrm{MOM}}(\mu)$  & $R^T(2\ \mathrm{GeV},\mu)$ \\
\hline
2 & 0.9676(13) & 0.9686(13) & - \\
3 & 0.97773(68) & 0.97934(71) & 1.03974(94) \\
4 & 0.98212(47) & 0.98390(48) & 1.0636(14) \\
\hline \hline
\end{tabular}
\end{table}

\begin{table*}[ht!]
\centering
\caption{Parameters of the MILC $n_{\mathrm{f}}=2+1+1$ HISQ gluon field ensembles we use.
Tensor current renormalisation factors in the RI$'$-MOM and RI-SMOM schemes 
are calculated on a subset of these ensembles: sets 1, 3, 5, 7 and 8 indicated by a $*$.
Labels for these configurations are given in the second column.
The third colum gives $\beta$: the bare QCD coupling for a wider range of ensembles.
The $J/\psi$ tensor decay constant is calculated on all of these 
ensembles. Two values of the lattice spacing are given, both in units of the 
Wilson flow parameter, $w_0$~\cite{Borsanyi:2012zs}. The physical value of 
$w_0$ is 0.1715(9) fm, fixed from $f_{\pi}$~\cite{fkpi}. Those denoted $a$ 
are calculated on each ensemble, and are the values used for the tensor decay 
constant. Those denoted $\tilde{a}$ 
are determined in the limit of physical sea masses at each value of 
$\beta$~\cite{Chakraborty:2014aca,Hatton:2020qhk}. This is the definition used in our 
calculation of the renormalisation factor, $Z_T$. Both determinations of the 
lattice spacing agree at the physical point. 
}
\label{tab:ensembles-smom}
\begin{tabular}{lllllllllll}
\hline \hline
Set & Label & $\beta$ & $w_0/a$ & $w_0/\tilde{a}$ & $L_s$ & $L_t$ & $am_l^{\mathrm{sea}}$ & $am_s^{\mathrm{sea}}$ & $am_c^{\mathrm{sea}}$ & $am_c^{\mathrm{val}}$ \\
\hline
1$*$ & very-coarse (vc) & 5.80 & 1.1272(7) & 1.1265(31) & 24 & 48 & 0.0064 & 0.064 & 0.828 & 0.873 \\
2 & - & 6.00 & 1.3826(11) & 1.4055(33) & 24 & 64 & 0.0102 & 0.0509 & 0.635 & 0.664 \\
3$*$ & coarse (c) & 6.00 & 1.4029(9) & 1.4055(33) & 32 & 64 & 0.00507 & 0.0507 & 0.628 & 0.650 \\
4 & - & 6.00 & 1.4116(9) & 1.4055(33) & 48 & 64 & 0.001907 & 0.05252 & 0.6382 & 0.643 \\
5$*$ & fine (f) & 6.30 & 1.9330(20) & 1.9484(33) & 48 & 96 & 0.00363 & 0.0363 & 0.430 & 0.439 \\
6 & - & 6.30 & 1.9518(7) & 1.9484(33) & 64 & 96 & 0.00120 & 0.0363 & 0.432 & 0.433 \\
7$*$ & superfine (sf) & 6.72 & 2.8960(60) & 3.0130(56) & 48 & 144 & 0.0048 & 0.024 & 0.286 & 0.274 \\
8$*$ & ultrafine (uf) & 7.00 & 3.892(12) & 3.972(19) & 64 & 192 & 0.00316 & 0.0158 & 0.188 & 0.194 \\
\hline \hline
\end{tabular}
\end{table*}

\begin{table*}
\caption{$Z_T^{\mathrm{SMOM}}$ values on the ensembles in Table~\ref{tab:ensembles-smom}
at different $\mu$ values along with the correlation matrices for these
different $\mu$ values on a given set. $Z_T^{\mathrm{SMOM}}$ converts the lattice tensor 
current into the SMOM scheme. }
\label{tab:ZTsraw}
\centering
\begin{tabular}{lllll}
\hline \hline
Set & $\mu = 2$ GeV & $\mu = 3$ GeV & $\mu = 4$ GeV & correlation \\
\hline
very-coarse (vc) & 1.07293(18) & - & - & - \\
coarse (c) & 1.10035(28) & 1.036117(92) & - & $\left(\begin{array}{ll} 1 & 0.824 \\ 0.824 & 1 \end{array}\right)$ \\
fine (f) & 1.13250(14) & 1.064991(56) & 1.030967(30) & $\left(\begin{array}{lll} 1 & 0.560 & 0.375 \\ 0.560 & 1 & 0.861 \\ 0.375 & 0.861 & 1 \end{array}\right)$ \\
superfine (sf) & 1.16641(40) & 1.09808(12) & 1.061844(57) & $\left(\begin{array}{lll} 1 & 0.828 & 0.866 \\ 0.828 & 1 & 0.896 \\ 0.866 & 0.896 & 1 \end{array}\right)$ \\
 ultrafine (uf) & 1.1791(17) & 1.11629(64) & - & $\left(\begin{array}{ll} 1 & 0.925 \\ 0.925 & 1 \end{array}\right)$ \\
 \hline \hline
 \end{tabular}
 \end{table*}

 \begin{table*}
 \caption{RI$'$-MOM equivalents ($Z_T^{\text{MOM}}$) of the RI-SMOM values in Table~\ref{tab:ZTsraw}.}
 \label{tab:ZTsrawpmom}
 \centering
 \begin{tabular}{lllll}
 \hline \hline
 Set & $\mu = 2$ GeV & $\mu = 3$ GeV & $\mu = 4$ GeV & correlation \\
 \hline
 very-coarse & 1.08435(42)& - & - & - \\
 coarse & 1.10970(58) & 1.04631(16) & - & $\left(\begin{array}{ll} 1 & 0.637 \\ 0.637 & 1 \end{array}\right)$ \\
 fine & 1.13949(47) & 1.06979(13) & 1.037388(39) & $\left(\begin{array}{lll} 1 & 0.384 & 0.393 \\ 0.384 & 1 & 0.609 \\ 0.393 & 0.609 & 1 \end{array}\right)$ \\
 superfine & 1.17449(71) & 1.10045(25) & 1.063735(93) & $\left(\begin{array}{lll} 1 & 0.103 & 0.155 \\ 0.103 & 1 & 0.337 \\ 0.155 & 0.337 & 1 \end{array}\right)$ \\
  ultrafine & 1.1845(29) & 1.1181(14) & - & $\left(\begin{array}{ll} 1 & 0.234 \\ 0.234 & 1 \end{array}\right)$ \\
  \hline \hline
  \end{tabular}
  \end{table*}

The wavefunction renormalisation may be calculated using either the incoming ($p_1$) or
outgoing ($p_2$) quark propagators. In the RI-SMOM scheme \cite{Sturm:2009kb} the momenta appearing in Eq.~\eqref{eq:GT} 
satisfy the symmetric conditions $p_1-p_2=q$ and $p_1^2=p_2^2=q^2\equiv \mu^2$. 

The amputated tensor vertex function $\Lambda_T$ is calculated 
by dividing $G_T$ on either side by the quark propagators: $\Lambda_T = S^{-1}(p_2)G_TS^{-1}(p_1)$.
The tensor current renormalisation factor, $Z_T$, that converts the 
lattice current into one in the momentum-subtraction 
scheme may then be defined as
\begin{equation} \label{eq:ZT-def}
  \frac{Z_q}{Z_T} = \frac{1}{144} \mathrm{Tr}(\Lambda_T^{\mu\nu}\sigma_{\mu\nu}) .
\end{equation}

Renormalisation factors taking the lattice to the RI-SMOM scheme, $Z_T^{\mathrm{SMOM}}$, 
can be converted to the more conventional
choice of the $\overline{\mathrm{MS}}$ scheme through a calculation in 
continuum perturbative QCD of the SMOM-to-$\overline{\mathrm{MS}}$ matching.
For the tensor renormalisation this has now been performed to three loop order \cite{Almeida:2010ns,Kniehl:2020sgo}. 
The RI-SMOM to $\overline{\mathrm{MS}}$ matching factor is:
\begin{equation}
\begin{split}
    &Z_{T}^{\overline{\mathrm{MS}}/\mathrm{SMOM}}(\mu,n_f) = 1 - 0.21517295 \frac{\alpha_{\overline{\mathrm{MS}}}(\mu)}{4 \pi} \\
    &- (43.38395 - 4.103279 n_f) \left( \frac{\alpha_{\overline{\mathrm{MS}}}(\mu)}{4 \pi} \right)^2 \\
    &- (1950.76(11) - 309.8285(28) n_f \\ &+ 7.063585(58) n_f^2) \left( \frac{\alpha_{\overline{\mathrm{MS}}}(\mu)}{4 \pi} \right)^3 .
\end{split}
\end{equation}
Evaluating this expression for $n_f = 4$ gives:
\begin{eqnarray} \label{eq:SMOM-matching}
    Z_T^{\overline{\mathrm{MS}}/\mathrm{SMOM}}(\mu) &=& 1 - 0.0171229 \alpha_{\overline{\mathrm{MS}}}(\mu)  \\
    &-& 0.170795 \alpha_{\overline{\mathrm{MS}}}^2(\mu) - 0.415470(55) \alpha_{\overline{\mathrm{MS}}}^3(\mu) . \nonumber
\end{eqnarray}

We also compare to results in the RI$'$-MOM scheme which has a simpler
kinematic setup than the RI-SMOM scheme. No momentum is inserted at the vertex
and therefore there is only one quark momentum, i.e. $p_1=p_2$, $q=0$. 
RI$'$-MOM uses the same definitions of $Z_q$
and $Z_T$ in Eq.~\eqref{eq:Zq-def} and Eq.~\eqref{eq:ZT-def}. 
The RI$'$-MOM to $\overline{\mathrm{MS}}$ conversion is also known through
$\mathcal{O}(\alpha_s^3)$ for the tensor current renormalisation factor~\cite{Gracey:2003yr}.
For $n_f=4$ the expression is:
\begin{equation} \label{eq:primeconv}
\begin{split}
    Z_T^{\overline{\mathrm{MS}}/\mathrm{MOM}}(\mu) &= 1 - 0.1976305 \alpha_{\overline{\mathrm{MS}}}^2(\mu) \\
    &- 0.4768793 \alpha_{\overline{\mathrm{MS}}}^3(\mu) .
\end{split}
\end{equation}
This is very similar to the RI-SMOM to $\overline{\mathrm{MS}}$ matching in Eq.~\eqref{eq:SMOM-matching} 
although with no $\mathcal{O}(\alpha_s)$ term in Landau gauge.
The situation is then very different from the case for the mass renormalisation factor where the 
RI-SMOM matching is considerably more convergent than the corresponding RI$'$-MOM 
matching~\cite{Franco:1998bm,Gracey:2003yr,Chetyrkin:2009kh,Sturm:2009kb,Gorbahn:2010bf,Almeida:2010ns}.

We tabulate the values of $Z_T^{\overline{\mathrm{MS}}/\mathrm{SMOM}}$ and 
$Z_T^{\overline{\mathrm{MS}}/\mathrm{MOM}}$ in columns 2 and 3 of Table~\ref{tab:convs} for 
different $\mu$ values. We also give the values required to run the tensor 
renormalisation factors in the $\overline{\mathrm{MS}}$ scheme to a reference scale of 
2 GeV, denoted $R^T(2\ \mathrm{GeV},\mu)$. These numbers are calculated using the three-loop tensor anomalous 
dimension \cite{Gracey:2000am}.

The work of \cite{Bi:2017ybi} compares RI$'$-MOM and RI-SMOM renormalisation for various 
currents. In the discussion of the tensor current presented there, uncertainties associated 
with missing terms in the matching to the $\overline{\mathrm{MS}}$ scheme were added to 
the renormalisation factors. As \cite{Bi:2017ybi} predates the results of \cite{Kniehl:2020sgo} 
a larger uncertainty was included on the RI-SMOM tensor renormalisation result of 
\cite{Bi:2017ybi} than on the RI$'$-MOM result. As both $\overline{\mathrm{MS}}$ 
conversion factors are now known to the same order in perturbation theory this issue 
has been removed for the comparison between the scheme. 
In Section~\ref{sec:jpsi-tensor} we address the issue of remaining 
uncertainty from unknown higher order terms in the conversion factors through 
our fits.

\section{Lattice calculation of $Z_T^{\mathrm{SMOM}}$ and $Z_T^{\mathrm{MOM}}$} \label{sec:ZT-latt}

We use the Highly Improved Staggered Quark (HISQ) action for both valence and sea quarks. The use of 
staggered quarks with momentum-subtraction schemes requires some consideration as 
explained in~\cite{Lytle:2013qoa}.
As discussed there, we take physical momenta to lie in the reduced Brillouin zone 
$-\pi/2 \leq p_{\mu}' \leq \pi/2$ and use momentum-space staggered quark fields 
at momenta $p' +\pi A$ where $A$ is a hypercubic vector of 1s and 0s. 
This multiplicity in momentum-space fields for a given physical momentum 
contains the staggered quark taste information. For each of these momenta we numerically solve 
the Dirac equation with a `momentum' source: $MS = e^{ip \cdot x}$ where $M$ is the Dirac matrix. 
This yields a quark propagator that we denote $S(p,x)$. The
gauge fields used in the construction of the Dirac matrix are numerically fixed
to Landau gauge by maximising the colour trace of the average link.

With the staggered quark fields $\chi$ the local tensor ($(\gamma_{\mu}\gamma_{\nu} \otimes \xi_{\mu}\xi_{\nu})$ 
in spin-taste notation) vertex function is
\begin{equation} \label{eq:vert-func}
  \begin{split}
  &\bigg \langle \chi(p_1' + \pi A) \\ &\left(\sum_x \overline{\chi}(x)(-1)^{(x_{\mu}+x_{\nu})}\chi(x)e^{i(p_1'-p_2')x} \right) \\ &\overline{\chi}(p_2' + \pi B) \bigg \rangle \\
  &= \frac{1}{n_{\mathrm{cfg}}} \sum_{x,\mathrm{cfg}} S(p_1' + \pi A,x)e^{i(p_1'-p_2')x} \times \\ &(-1)^{x_0+x_1+x_2+x_3-x_{\mu}-x_{\nu}} S^{\dagger}(p_2' + \pi \tilde{B},x) ,
  \end{split}
\end{equation}
making use of the $\gamma_5$-hermiticity of $S$ in the last line. 
The elements of $\tilde{B}$ are permuted compared to
those of $B$ via $\tilde{B} = B +_2 (1,1,1,1)$ where $+_2$ denotes addition modulo 2.

We use the following kinematic setup, which obeys the symmetric conditions of 
the RI-SMOM scheme:
\begin{equation}
\begin{split}
    ap_1' &= \frac{2\pi}{L_s} \left( x+\frac{\theta}{2},0,x+\frac{\theta}{2},0 \right) , \\
    ap_2' &= \frac{2\pi}{L_s} \left( x+\frac{\theta}{2},-x-\frac{\theta}{2},0,0 \right) , \\
    aq' &= \frac{2\pi}{L_s} \left( 0,x+\frac{\theta}{2},x+\frac{\theta}{2},0 \right) .
\end{split}
\end{equation}
$x$ is an integer and $\theta$ is the momentum-twist applied with phased 
boundary conditions that we use to access arbitrary momenta~\cite{Arthur:2010ht}.
For the single momentum in the RI$'$-MOM scheme we use $ap_1'$.

Our calculations are done on HISQ $n_f = 2+1+1$ gluon field ensembles 
generated by the MILC collaboration~\cite{Bazavov:2010ru,Bazavov:2012xda},   
the details of which are given in Table~\ref{tab:ensembles-smom}.
On each ensemble we use 20 configurations except
for ultrafine where only 6 configurations with stringent gauge fixing were
available. We have checked, using other sets, that 
this small number of configurations is sufficient to achieve
high precision given our use of momentum sources.
In order to compensate for a potential underestimation of the
uncertainty from the low statistics, however, 
we double the uncertainty on the $Z_T$
values on set 8.

Table~\ref{tab:ensembles-smom} gives two values for the lattice spacing, reflecting the 
different approach to the physical quark mass limit that we take in the two parts 
of our calculation. Both approaches arrive at the same physical point, so this is simply 
a convenient choice away from the physical point. We label the two lattice spacing values 
$a$ and $\tilde{a}$.  
$a$ is determined from a calculaton of $w_0/a$~\cite{Borsanyi:2012zs} on each 
ensemble and varies as the sea quark masses are changed at fixed bare gauge coupling, $\beta$.
$\tilde{a}$ is the value of the lattice spacing for physical 
sea quark masses at a given value of $\beta$~\cite{Chakraborty:2014aca,Hatton:2020qhk}. 
The latter definition is used for the calculation of $Z_T$ while 
the former is used to compute the $J/\psi$ tensor decay constant.

We use different definitions of the lattice spacing to reduce the effects of sea quark mass 
mistuning in the calculation. If we instead used a single definition of the 
lattice spacing we would 
have a steeper approach to the tuned sea quark mass point either in the renormalisation factors or in the hadronic matrix elements.  
Hadronic matrix elements are sensitive to low energy scales and it is convenient to 
keep the value of $w_0$ fixed as the sea quark masses are varied, leading to values of $w_0/a$ that 
are dependent on the sea quark masses. As discussed in Appendix A of~\cite{Chakraborty:2014aca} 
the variation of hadronic quantities with the sea quark masses is similar to that of $w_0$ and so 
they do not vary much if $w_0$ is held fixed. Sea quark mass dependence in the hadronic quantity in 
lattice units is cancelled by the variation of $w_0/a$. However, ultraviolet quantities such as renormalisation 
factors have very weak sea quark mass dependence. Using $w_0/a$ values that vary with the sea
masses therefore introduces unwanted dependence and so we choose to 
use $w_0/\tilde{a}$ defined in the physical sea quark mass limit. 
The sea quark mass dependence of RI-SMOM renormalisation factors 
was studied in~\cite{Lytle:2018evc} using $w_0/\tilde{a}$ and indeed found to be tiny.
We will see from the plots of our results in the next Section that our strategy 
of using $a$ and $\tilde{a}$
does indeed lead to very little difference between results for 
physical and unphysical sea quark masses for 
the decay constant. 

\begin{figure}
\centering
\includegraphics[width=0.45\textwidth]{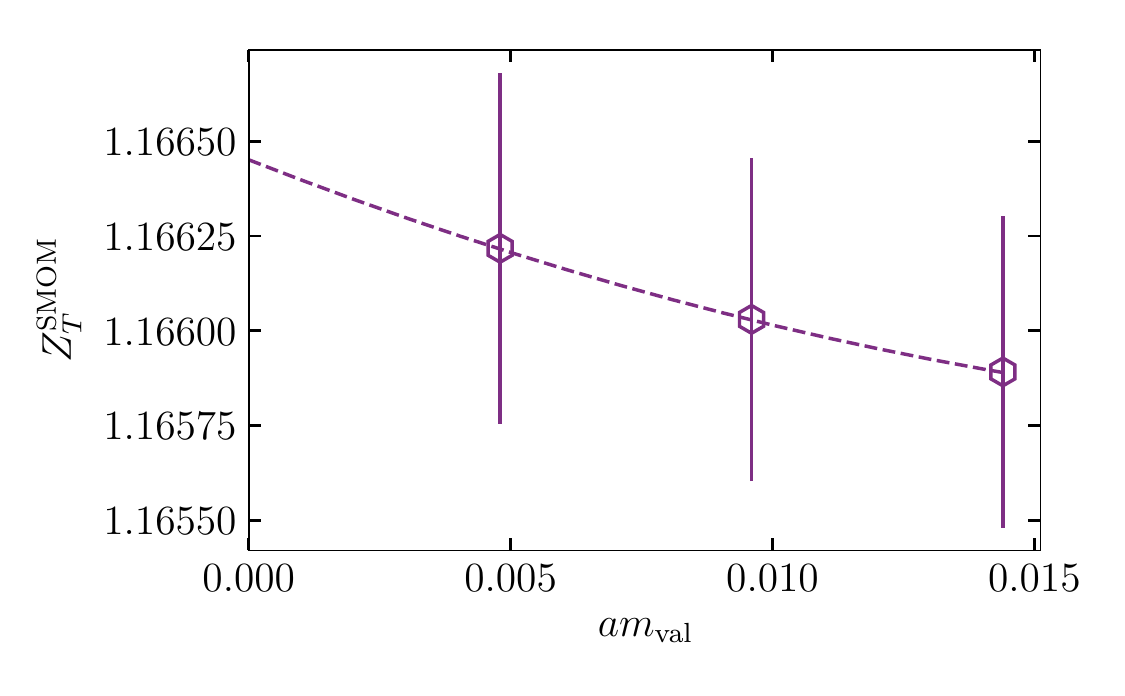}
\includegraphics[width=0.45\textwidth]{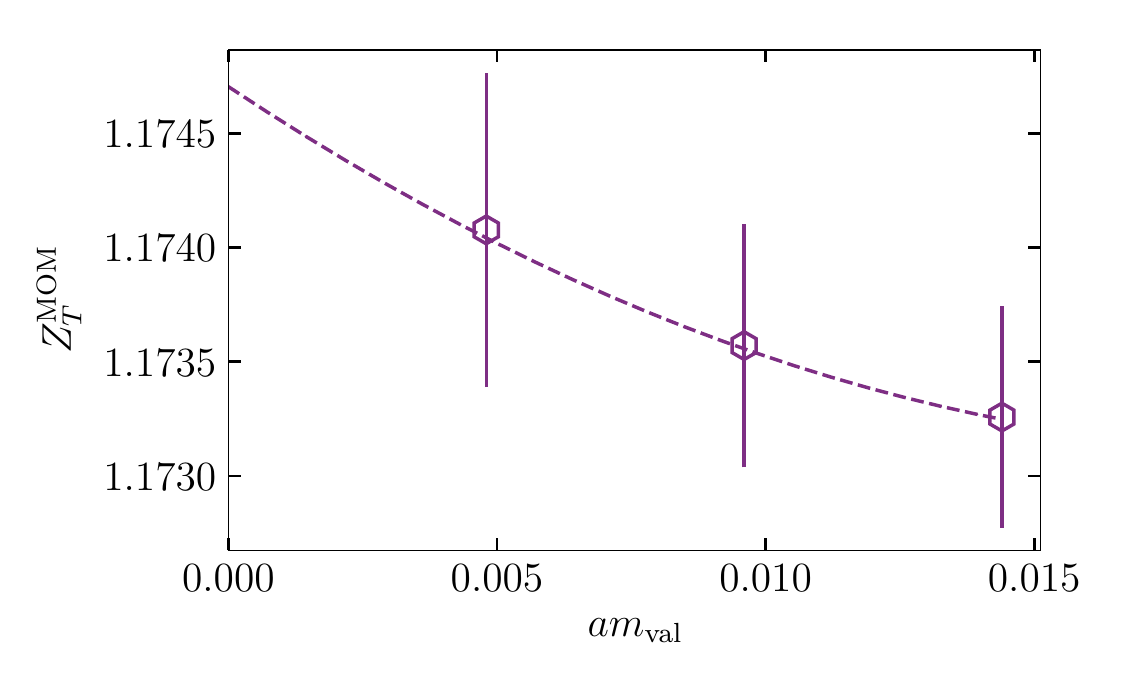}
\caption{Valence mass dependence of tensor current renormalisation factors in
the RI-SMOM (upper) and RI$'$-MOM (lower) schemes. The values shown are for
$\mu = 2$ GeV on Set 7. Both show reasonably mild dependence on the valence
mass but the dependence is smaller for RI-SMOM.}
\label{fig:ZT-mass-dep}
\end{figure}

We define the RI-SMOM and RI$'$-MOM schemes at zero valence quark mass 
to remove mass-dependent non-perturbative contributions.
In order to obtain values at zero valence mass we calculate
$Z_T$ at three different quark masses and extrapolate to 0 using a polynomial
fit in $am_{\mathrm{val}}$:
\begin{equation}
Z_T(am_{\mathrm{val}},\mu) = Z_T(\mu) + d_1(\mu)\frac{am_{\mathrm{val}}}{am_s} d_1(\mu)\left(\frac{am_{\mathrm{val}}}{am_s} \right)^2 .
\end{equation}
The three valence masses that we use are $\{am^{\mathrm{sea}}_l,2am^{\mathrm{sea}}_l,3am^{\mathrm{sea}}_l\}$.
This is the same procedure as was used in \cite{Lytle:2018evc} 
and \cite{Hatton:2019gha}. 
Fig.~\ref{fig:ZT-mass-dep} shows an example of the mass dependence of 
$Z_T$ for both the lattice-to-SMOM matching factor, $Z_T^{\mathrm{SMOM}}$, and 
the lattice-to-$'$MOM factor, $Z_T^{\mathrm{MOM}}$. The mass dependence reflects 
non-perturbative artefacts (condensates) appearing in $Z_T$ with mass-dependent 
coefficients. We see that the dependence is very small for the SMOM case and 
less so, but still relatively benign, in the MOM case. 

We collect our $Z_T^{\mathrm{SMOM}}$ results, extrapolated to zero valence mass, for various 
values of $\mu$ in Table~\ref{tab:ZTsraw}. The correlation matrix for these different 
$\mu$ values on each ensemble is also given. Our $Z_T^{\mathrm{MOM}}$ results are similarly 
collected in Table~\ref{tab:ZTsrawpmom}.

\section{$J/\psi$ tensor decay constant} \label{sec:jpsi-tensor}

The $J/\psi$ tensor decay constant, $f_{J/\psi}^T$, is defined in 
an analogous way to the 
$J/\psi$ vector decay constant $f_{J/\psi}^V$. 
$f_{J/\psi}^T$ 
parameterises the vacuum to meson matrix element of a tensor current in the following
way:
\begin{equation} \label{eq:mat-element}
  \langle 0 \vert \overline{\psi} \sigma_{\alpha\beta} \psi \vert J/\psi \rangle = i f_{J/\psi}^T(\mu) (\epsilon_{\alpha}p_{\beta} - \epsilon_{\beta}p_{\alpha}) .
\end{equation}
$\epsilon$ is the polarisation vector of the $J/\psi$, $p$ 
is the $J/\psi$ 4-momentum and $\mu$ is the renormalisation scale 
for the tensor decay constant. Note that the tensor decay constant
is $\mu$-dependent, reflecting the anomalous dimension of the 
continuum tensor current and unlike the vector decay constant. 
It is also scheme-dependent and we will give results in the 
$\overline{\text{MS}}$ scheme.

If one of the indices of 
the tensor current is in the time direction, we can 
extract $f_{J/\psi}^T$ from the 2-point tensor-tensor correlation function
projected onto zero spatial momentum. We construct this as
\begin{equation} \label{eq:correlator}
C_T (t) = \frac{1}{4} \sum_{x} \langle (-1)^{\eta_T(x)}\mathrm{Tr}(S(x,0)S^{\dagger}(x,0)) \rangle .
\end{equation}
Here $\eta_T(x)$ is a position-dependent phase remnant of $\sigma_{\alpha\beta}$
resulting from the use of staggered quarks. This is the same phase as that
appearing in Eq.~\eqref{eq:vert-func}, since we use the same local tensor current. 
We take $\beta$ to be in the temporal
direction and average $\alpha$ over spatial directions. 

We compute the correlation function of Eq.~\eqref{eq:correlator} on the full 
set of
ensembles with parameters summarised in Table~\ref{tab:ensembles-smom}. 
The valence $c$ quark masses are chosen to be close to those giving the 
experimental value of the $J/\psi$ mass~\cite{Hatton:2020qhk}. We will allow 
for mistuning of the valence $c$ quark mass in our fits to extrapolate to 
the continuum limit. 
The decay constant is determined from the ground-state parameters 
extracted from a multi-exponential fit to the averaged 2-point correlator: 
\begin{eqnarray}
\langle C_T(t)\rangle &=& \sum_i \left( A^T_i f(E_i^T,t) - (-1)^t A^{T,o}_i f(E^{T,o}_i ,t) \right) , \nonumber \\
f(E,t) &=& e^{-Et} + e^{-E(L_t-t)} .
\end{eqnarray}
The temporal oscillation term appears because of our use of staggered quarks.
We perform the fit 
using standard Bayesian fitting techniques~\cite{Lepage:2001ym} with broad priors on the parameters, 
as in~\cite{Hatton:2020qhk}.  

The $J/\psi$ tensor decay constant is then calculated from the ground-state amplitude 
and energy according to 
\begin{equation}
f_{J/\psi}^T = Z_T \sqrt{\frac{2A_0^T}{E_0^T}} .
\end{equation}
Here the ground state energy, $E_0^T$, is the mass of the $J/\psi$ as 
we implement Eq.~\eqref{eq:mat-element} for a $J/\psi$ at rest.  

\begin{figure}
\centering
\includegraphics[width=0.45\textwidth]{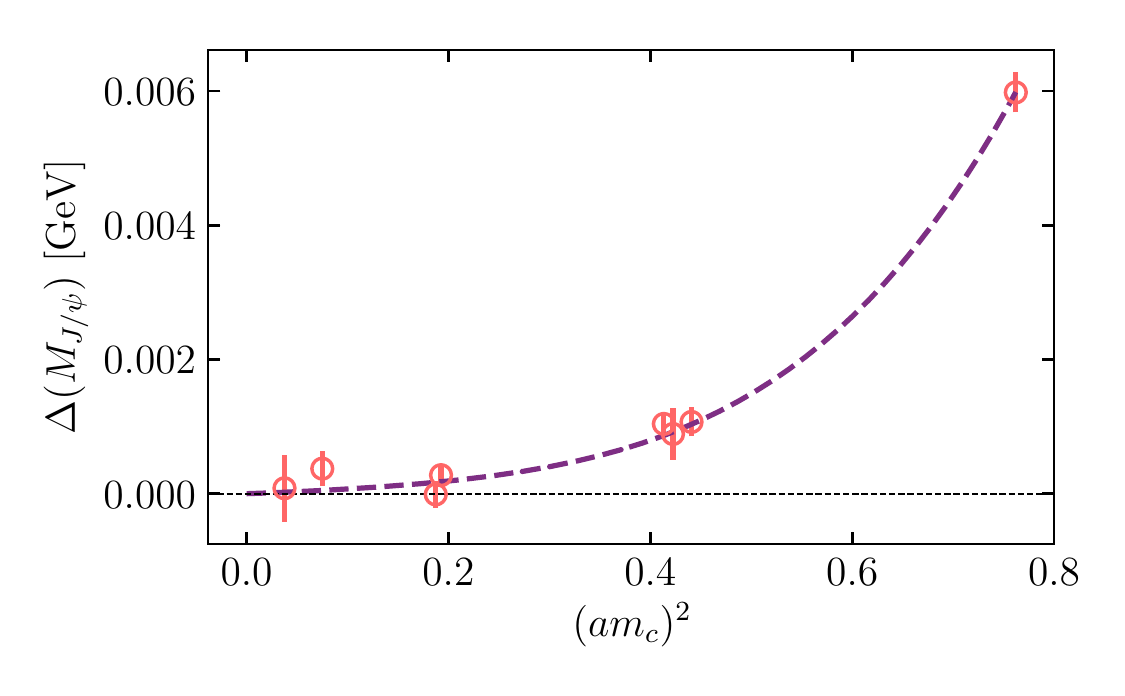}
\caption{The taste splitting of the $J/\psi$ masses of tastes $\xi_{\mu}\xi_t$ and 
$\xi_{\mu}$ as a function of $(am_c)^2$. Clearly this difference decreases with the 
lattice spacing. The fit line shown has the form of Eq.~\eqref{eq:deltaM-fit}.}
\label{fig:jpsi-masses}
\end{figure}

As we have used 
the local tensor current with taste $\xi_{\alpha}\xi_t$, $E_0^T$ is the mass of 
the $J/\psi$ of that taste. Because of taste splitting effects this is expected to 
differ from the local $J/\psi$ with taste $\xi_{\alpha}$. The values of the local 
$J/\psi$ mass on the ensembles used here were given in \cite{Hatton:2020qhk} and 
we collect the values for the taste $\xi_{\alpha}\xi_t$ in Table~\ref{tab:aft}.
As taste-breaking effects are a discretisation effect we should see the difference 
between the two masses ($\Delta(M_{J/\psi})$) decrease as the lattice spacing is 
decreased. This is shown in Fig.~\ref{fig:jpsi-masses}. Note that even on the coarsest 
ensemble the difference is only 6 MeV, about 0.2\% of the $J/\psi$ mass. 
A fit to the mass difference 
of the form 
\begin{equation} \label{eq:deltaM-fit}
\Delta(M_{J/\psi})(a) = c_1 \alpha_s(1/a) (am_c)^2 + c_2 (am_c)^4 ,
\end{equation}
is included in the figure. This is the expected form for taste effects 
as the HISQ action is improved to remove 
tree-level $(am_c)^2$ errors~\cite{Follana:2006rc}. 
The fit works well, with a $\chi^2/\mathrm{dof}$ of 0.4.

The values of $af_{J/\psi}^T/Z_T$ extracted from our 2-point correlator fits on
the ensembles in Table~\ref{tab:ensembles-smom} are given in Table~\ref{tab:aft}.

\begin{table}
\centering
\caption{Results for the $J/\psi$ tensor decay constant on each of the ensembles in
Table~\ref{tab:ensembles-smom} in lattice units before multiplication by the
tensor renormalisation factor. We also give the ratio of the tensor and vector $J/\psi$
decay constants (again, before renormalisation) in column 3. We give the $J/\psi$ mass 
extracted from our 2-point correlator fits in column 4. This is the mass for a $J/\psi$ of taste 
$\xi_{\alpha}\xi_t$.}
\label{tab:aft}
\begin{tabular}{llll}
\hline \hline
Set & $af_{J/\psi}^T / Z_T$ & $(Z_V f_{J/\psi}^T)/(Z_T f^V_{J/\psi})$ & $aM_{J/\psi}$ \\
\hline
1 & 0.3741(12) & 0.8837(30) & 2.39769(18) \\
2 & 0.25754(15) & 0.87548(81) & 1.944312(92) \\
3 & 0.25212(35) & 0.8743(13) & 1.91530(23) \\
4 & 0.24977(36) & 0.8747(13) & 1.901880(40) \\
5 & 0.165404(96) & 0.86433(62) & 1.391514(65) \\
6 & 0.16396(13) & 0.86386(78) & 1.378232(73) \\
7 & 0.105293(93) & 0.8535(10) & 0.929972(57) \\
8 & 0.07685(19) & 0.8410(22) & 0.691999(97) \\
\hline \hline
\end{tabular}
\end{table}

An important goal of this analysis is to investigate the size of
systematic effects arising from nonperturbative contamination of $Z_T$ and show how to remove
them. Doing this requires analysis of a physical quantity sensitive to the tensor 
current renormalisation, 
for which we use the $J/\psi$ tensor decay
constant in the $\overline{\mathrm{MS}}$ scheme at a reference scale 
of 2 GeV. This is obtained by taking the product of several quantities: 
the unrenormalised
$J/\psi$ tensor decay constant $af_{J/\psi}^T/Z_T$ from Table~\ref{tab:aft}; 
the renormalisation factor that converts this to a momentum-subtraction scheme at scale $\mu$ from 
Table~\ref{tab:ZTsraw} or Table~\ref{tab:ZTsrawpmom} (although for convenience here we use 
SMOM notation); the perturbative matching from the momentum-subtraction scheme to 
$\overline{\text{MS}}$ (discussed in Section~\ref{sec:ZT-intro}) 
and the running from $\mu$ to 2 GeV in the $\overline{\mathrm{MS}}$ scheme.
These last two factors are given in Table~\ref{tab:convs}. This gives us the 
results that we will fit:
\begin{equation} \label{eq:fT-construct}
\begin{split}
    f_{J/\psi}^T(\overline{\text{MS}},2\ \mathrm{GeV},\mu,a) &= R^T(2\ \mathrm{GeV},\mu) Z_T^{\overline{\mathrm{MS}}/\mathrm{SMOM}}(\mu) \\ &\times Z_T^{\mathrm{SMOM}}(\mu,a) (af_{J/\psi}^T/Z_T)/a .
\end{split}
\end{equation}
Note that the first three factors above, combined, constitute 
$Z_T^{\overline{\text{MS}}}(2\ \mathrm{GeV},a)$ i.e. the renormalisation factor that takes the decay constant from the lattice scheme 
to the $\overline{\text{MS}}$ scheme at a renormalisation scale of 2 GeV, up to discretisation 
effects and nonperturbative artefacts present in $Z_T^{\mathrm{SMOM}}$. 

We fit the results from Eq.~\ref{eq:fT-construct} as a function of lattice 
spacing and $\mu$ values in order to obtain 
a physical value for $f_{J/\psi}^T(\overline{\text{MS}},2\ \mathrm{GeV})$ 
in the continuum limit. The fit form used is:
\begin{equation} \label{eq:fit-form}
\begin{split}
f_{J/\psi}^T (\overline{\text{MS}},\mu_{\mathrm{ref}},\mu,a) &= f_{J/\psi}^T (\overline{\text{MS}},\mu_{\mathrm{ref}})
\times \\ 
&\hspace{-8.0em} \bigg[1 + \sum_{n} c_{am_c}^{(n)} (am_c)^{2n} 
+ h_{\ell}^{\mathrm{sea}} \frac{\delta_{m_{\ell}}^{\mathrm{sea}}}{m_{s}^{\mathrm{phys}}}
+ h_{c}^{\mathrm{sea}} \frac{\delta_{m_{c}}^{\mathrm{sea}}}{m_{c}^{\mathrm{phys}}} \\ &\hspace{2.0em} + h_{c}^{\mathrm{val}} \frac{M_{J/\psi} - M_{J/\psi}^{\mathrm{expt}}}{M_{J/\psi}^{\mathrm{expt}}} \bigg] \times \\
&\hspace{-8.0em}\bigg [1 + \sum_i c_{a\mu}^{(i)} (\tilde{a}\mu/\pi)^{2i} +
\alpha_{\overline{\mathrm{MS}}}^4 (\mu) \left(c_{\alpha 1} +
c_{\alpha 2} \log(\mu/\mu_{\text{ref}})\right)\\
&+ \sum_j c_{\mathrm{cond}}^{(j)} \alpha_{\overline{\mathrm{MS}}}(\mu)
\frac{(1\ \mathrm{GeV})^{2j}}{\mu^{2j}} \bigg] .
\end{split}
\end{equation}
This is designed to capture
the lattice spacing and $\mu$ dependence of $Z_T$ as well as the discretisation and 
quark mass effects in $af^T_{J/\psi}/Z_T$. We take $\mu_{\mathrm{ref}}$ to be 2 GeV and 
include results from $\mu$ values of 2, 3 and 4 GeV and multiple values of $a$.

The first square brackets of Eq.~(\ref{eq:fit-form}) allow for discretisation effects 
in the raw lattice values for $af^T_{J/\psi}$ through an even polynomial in powers of 
the $c$ quark mass in lattice units, $am_c$, as appropriate for a charmonium quantity. 
The next terms in that bracket then account for mistuning of the sea quark masses away 
from their physical values and mistuning of the valence $c$ quark mass, respectively. 
This part of the fit is the same form as that used 
for the $J/\psi$ vector decay constant in~\cite{Hatton:2020qhk}. 

The second set of square brackets in Eq.~(\ref{eq:fit-form}) allows for effects 
from the lattice calculation of $Z_T$ in the momentum-subtraction scheme at scale 
$\mu$. We expect discretisation effects in this case to appear as even 
powers of $\tilde{a}\mu/\pi$. The missing $\alpha_s^4$ term in the matching from 
momentum-subtraction to $\overline{\text{MS}}$ schemes is allowed for with 
coefficient $c_{\alpha 1}$ and a similar effect for the running, with 
coefficient $c_{\alpha 2}$. The terms on the final line allow for the 
condensate contamination of $Z_T$ coming from its nonperturbative calculation 
on the lattice. The condensate contamination is visible in an Operator 
Product Expansion of, for example, the quark propagator~\cite{Chetyrkin:2009kh} 
where it appears in terms suppressed by powers of the renormalisation scale 
$\mu$. For the gauge-fixed quantities that we calculate here to determine $Z_T$ 
these terms appear first at $\mathcal{O}(1/\mu^2)$ multiplied by the Landau
gauge gluon condensate $\langle A^2 \rangle$~\cite{Lytle:2018evc}. 
We also allow for higher order condensates 
with larger inverse powers of $\mu$, up to and including $1/\mu^6$.  

We take priors on all the coefficients of the fit in Eq.~(\ref{eq:fit-form})
of $0\pm 1$, except for three terms. We take a prior of $0\pm 0.1$ for $h_c^{\mathrm{sea}}$
based on~\cite{Hatton:2020qhk}, and $0\pm 0.5$ for $c_{\alpha 1}$
and $0\pm 0.4$ for $c_{\alpha 2}$
based on the lower order terms in Eqs.~(\ref{eq:SMOM-matching}) 
and~(\ref{eq:primeconv}) and in~\cite{Gracey:2000am}. We also take 
$0.4\pm 0.1$ GeV for the prior for the physical 
value of $f^T_{J/\psi}(\overline{\text{MS}}, 2\,\text{GeV})$ 
based on the expectation that it should be close
in value to $f^V_{J/\psi}$.  
We include 5 terms in each of the sums over discretisation effects and 3 terms in 
the sum over condensate contributions. 

\begin{figure}
\centering
\includegraphics[width=0.45\textwidth]{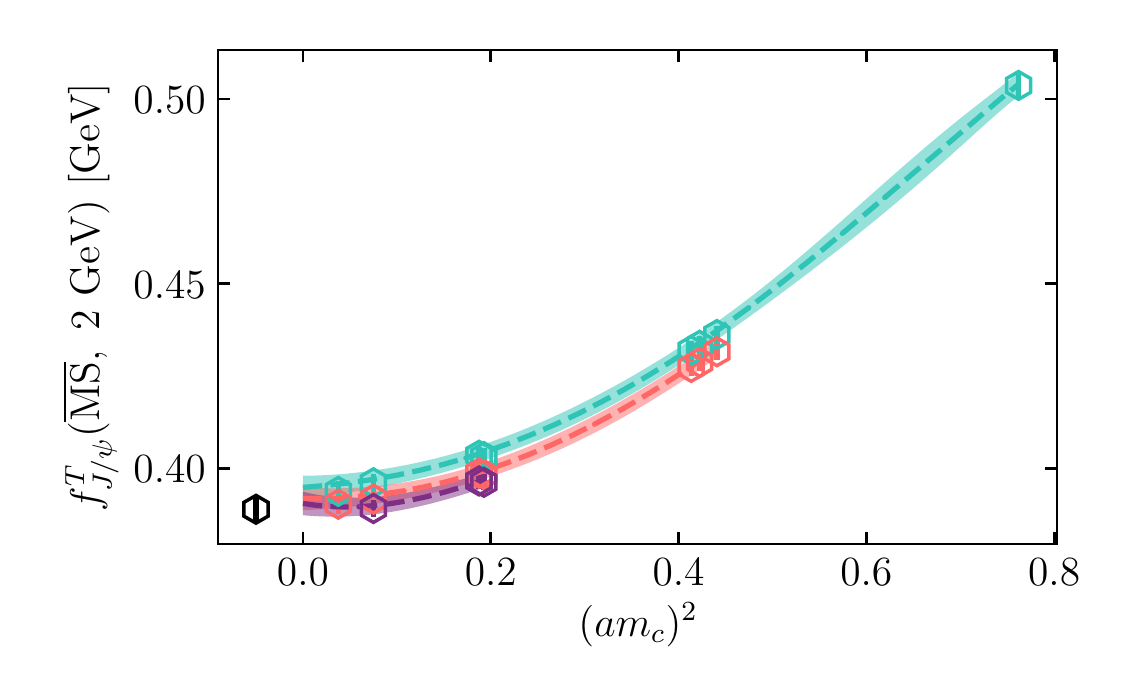}
\caption{Continuum extrapolation of the $J/\psi$ tensor decay constant in the
    $\overline{\mathrm{MS}}$ scheme at a scale of 2 GeV using
    lattice tensor current renormalisation in the RI-SMOM scheme at multiple $\mu$ values. Three different values of
    the renormalisation scale $\mu$ are used in the lattice calculation of $Z_T^{\mathrm{SMOM}}$
    to allow nonperturbative $\mu$
    dependence to be fitted. These three different $\mu$ values are shown as
    different coloured lines. The blue line is 2 GeV, the orange, 3 GeV and the purple,
    4 GeV. The black hexagon is the physical result for $f_{J/\psi}^T(2\ \mathrm{GeV})$
    obtained from the fit of Eq.~\eqref{eq:fit-form} (with the condensate pieces removed).}
\label{fig:fjpsit}
\end{figure}

Our results using the RI-SMOM $Z_T$ from Section~\ref{sec:ZT-latt} with the fit of Eq.~\eqref{eq:fit-form} are shown
in Fig.~\ref{fig:fjpsit}. The $\chi^2/\mathrm{dof}$ is 0.19 giving a continuum value with condensate contributions
from $Z_T$ removed of:
\begin{equation}
f^T_{J/\psi}(\overline{\text{MS}},2\ \mathrm{GeV}) = 0.3889(33)\ \mathrm{GeV}\ \ (\mathrm{int.\,SMOM}). 
\end{equation}
The phrase `int. SMOM' here indicates that the result uses the intermediate RI-SMOM scheme. 
Note that the $\chi^2/\mathrm{dof}$ increases significantly, to 2.5, if the 
$\mu$-dependent terms that survive the continuum limit, that is 
condensate terms and $\alpha_s^4$ terms, are removed from the fit. 

The black hexagon in Figure~\ref{fig:fjpsit} shows this 
result (the $f_{J/\psi}^T(\overline{\text{MS}},2\,\text{GeV})$ fit parameter in
Eq.~\eqref{eq:fit-form}). This is the physical value of the tensor decay constant, 
with discretisation and quark mass-mistuning effects extrapolated away 
and condensate contributions and $\alpha_s^4$ errors removed.
Note that this value is lower than the value obtained from simply taking the continuum limit of the
2 GeV results (blue line), mainly because of condensate contamination at $\mu=2$ GeV. 
This underlines the necessity of performing the calculation at multiple
values of $\mu$ in the RI-SMOM scheme before running all of the results to a reference scale, in
this case 2 GeV, in order to determine and remove systematic $\mu$-dependent errors. 

\begin{table}
\centering
\caption{Values, with uncertainties, 
and correlation matrix for the correction $C^{\text{SMOM}}(\mu)$ 
to be applied to renormalisation factors for the tensor current when using the 
RI-SMOM intermediate scheme. 
}
\label{tab:csmom}
\begin{tabular}{lllll}
\hline \hline
$\mu$ (GeV) & $C^{\text{SMOM}}(\mu)$ & \multicolumn{3}{c}{correlation matrix}\\
\hline
2 & 0.0153(36) & 1.0 & 0.9889 & 0.9249 \\
3 & 0.0074(24) & 0.9889 & 1.0 & 0.9708 \\
4 & 0.0041(16) & 0.9249 & 0.9708 & 1.0 \\
\hline \hline
\end{tabular}
\end{table}

The difference between the black hexagon and the continuum limit of 
the lines for the different $\mu$
values can be thought of as a correction that needs to be applied to the $Z_T$ values 
that connect the lattice results and the $\overline{\text{MS}}$ value at 
2 GeV (i.e. $Z_T^{\overline{\text{MS}}}(2\,\text{GeV},a)$ that combines the first 
three factors on the righthand side of Eq.~\eqref{eq:fT-construct}) 
so that they are independent of $\mu$. 
This will give a corrected $Z_T^{\overline{\text{MS}}}$ 
that can then be used in future calculations.
The correction depends on the intermediate momentum-subtraction scheme used and the 
condensate contamination that it has as well as $\alpha_s^4$ errors in the matching 
to $\overline{\text{MS}}$. 

\begin{figure}
\centering
\includegraphics[width=0.45\textwidth]{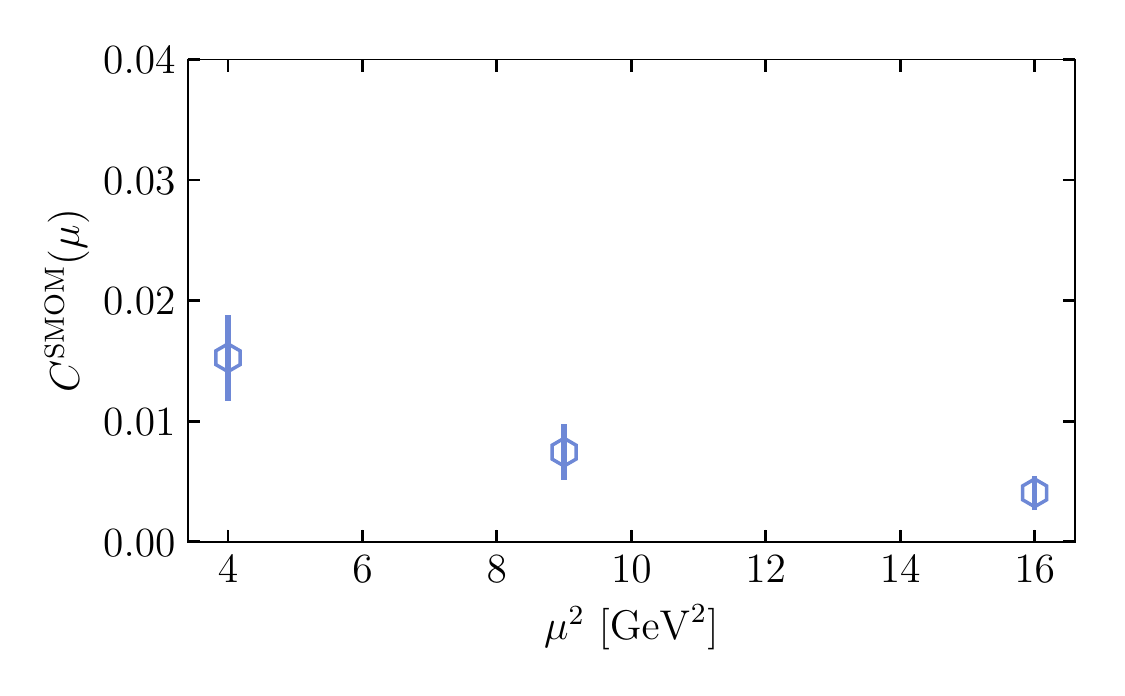}
\caption{The correction, $C^{\text{SMOM}}(\mu)$, to the tensor current renormalisation factor, 
$Z_T^{\overline{\text{MS}}} (2\,\text{GeV})$, required to
    account for nonperturbative effects arising from condensate contributions 
to the lattice calculation of $Z_T^{\text{SMOM}}(\mu)$ and missing $\alpha_s^4$ terms 
in the matching to $\overline{\text{MS}}$. The correction is defined in terms of a
    subset of the fit posteriors of the fit shown in Fig.~\ref{fig:fjpsit} (see text). 
}
\label{fig:correction}
\end{figure}

We define a $\mu$-dependent subtraction, $C^{\mathrm{SMOM}}(\mu)$, to apply to
the values of $Z_T^{\text{MS}}$ 
from the combination of the 
$c_{\mathrm{cond}}^{(j)}$ terms in Eq.~\eqref{eq:fit-form} along with the 
$c_{\alpha 1}$ and $c_{\alpha 2}$ terms. It is difficult for the fit 
to completely separate these
different $\mu$-dependent contributions and as a result the 
individual coefficients are not as well
determined as the total correction (because the fit parameters 
are correlated). 
The full correction is shown in
Fig.~\ref{fig:correction} plotted against $\mu^2$, and significantly non-zero 
values are seen across the $\mu^2$ range, with the correction
at the $\sim 1.5\%$ level for $\mu=2$ GeV. These values, and their 
correlation matrix, are given in Table~\ref{tab:csmom}. If we extract the condensate 
contributions to the correction separately, values with the same 
central values are obtained but with uncertainties that are about 
40\% larger at $\mu=2$ GeV. 
If the corrected $Z_T$ value is denoted 
$Z_T^{\overline{\text{MS}},\text{c}}$ and the uncorrected value 
$Z_T^{\overline{\text{MS}},\mathrm{u}}$, 
\begin{equation}
\label{eq:corrn}
Z_T^{\overline{\text{MS}},\mathrm{c}}(2\,\mathrm{GeV},a) = Z_T^{\overline{\text{MS}},\mathrm{u}}(2\,\text{GeV},a) - C^{\mathrm{SMOM}}(\mu) .
\end{equation}
A corrected value for $Z_T^{\overline{\text{MS}}}$ is then readily derived using the results in 
Tables~\ref{tab:ZTsraw},~\ref{tab:convs} and~\ref{tab:csmom}.

\begin{figure}
\centering
\includegraphics[width=0.45\textwidth]{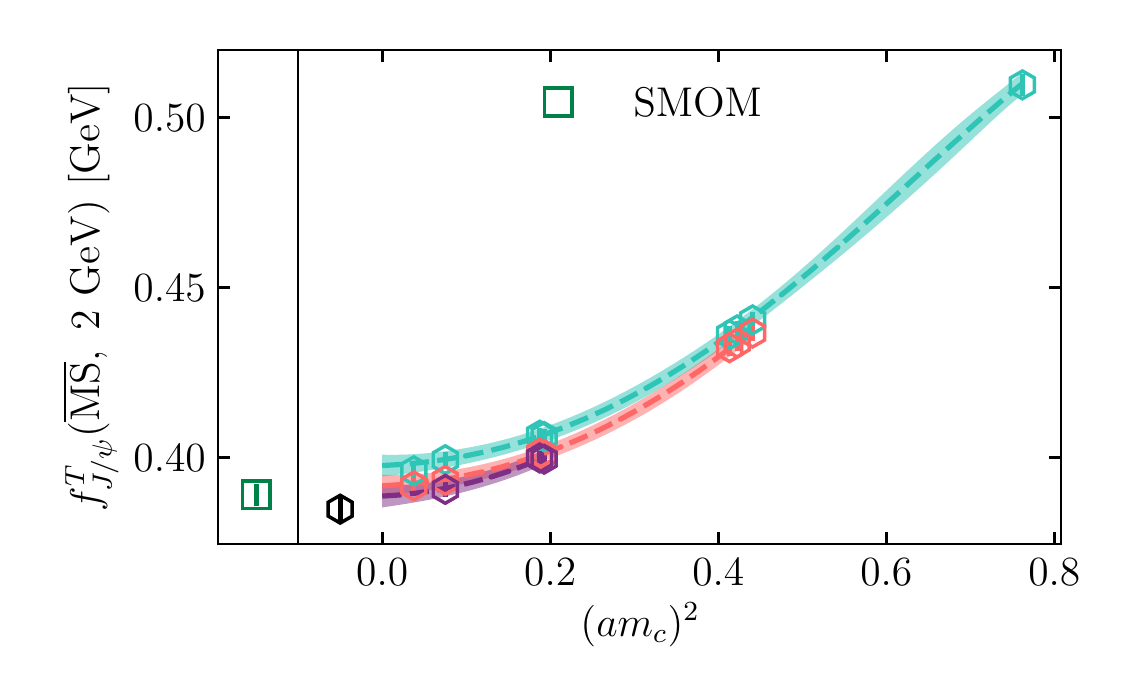}
\caption{Continuum extrapolation of the $J/\psi$ tensor decay constant in the $\overline{\text{MS}}$ 
scheme at a scale of 2 GeV using an intermediate
    nonperturbative renormalisation of the tensor current in the RI$'$-MOM scheme on the lattice.
    Multiple values of the renormalisation scale $\mu$ have been used so that
    $\mu$ dependent nonperturbative effects can be removed in the fit. 
    The blue points and line are for $\mu=2$ GeV, orange for 3 GeV and purple for 4 GeV. 
    The value obtained in the continuum limit with the condensate terms removed is shown 
    as a black hexagon.
The result
    is in agreement with that using RI-SMOM renormalisation (Figure~\ref{fig:fjpsit}) 
which is shown as the green square. 
}
\label{fig:fjpsitpm}
\end{figure}

We also examine $f_{J/\psi}^T$ using a tensor current renormalisation obtained
in the RI$'$-MOM scheme on the lattice. In this case we use the conversion to
$\overline{\mathrm{MS}}$ in Eq.~\eqref{eq:primeconv} and calculate the RI$'$-MOM 
equivalent of Eq.~\eqref{eq:fT-construct}. The results and the fit to Eq.~\eqref{eq:fit-form} 
are shown in
Fig.~\ref{fig:fjpsitpm}. We see that the final continuum result with condensate contributions 
and $\alpha_s^4$ errors removed agrees with that given by
intermediate RI-SMOM renormalisation factors. The $\chi^2/\mathrm{dof}$ of this fit is 0.4 giving a final
result of 
\begin{equation}
\label{eq:finvalpmom}
f_{J/\psi}^T(\overline{\text{MS}},2\ \mathrm{GeV}) = 0.3847(37)\ \mathrm{GeV}\ \ (\mathrm{int.\, MOM}). 
\end{equation}
Dropping both condensate and $\alpha_s^4$ terms from the fit increases the 
$\chi^2/\mathrm{dof}$ here to 8.2. 

\begin{figure}
\centering
\includegraphics[width=0.45\textwidth]{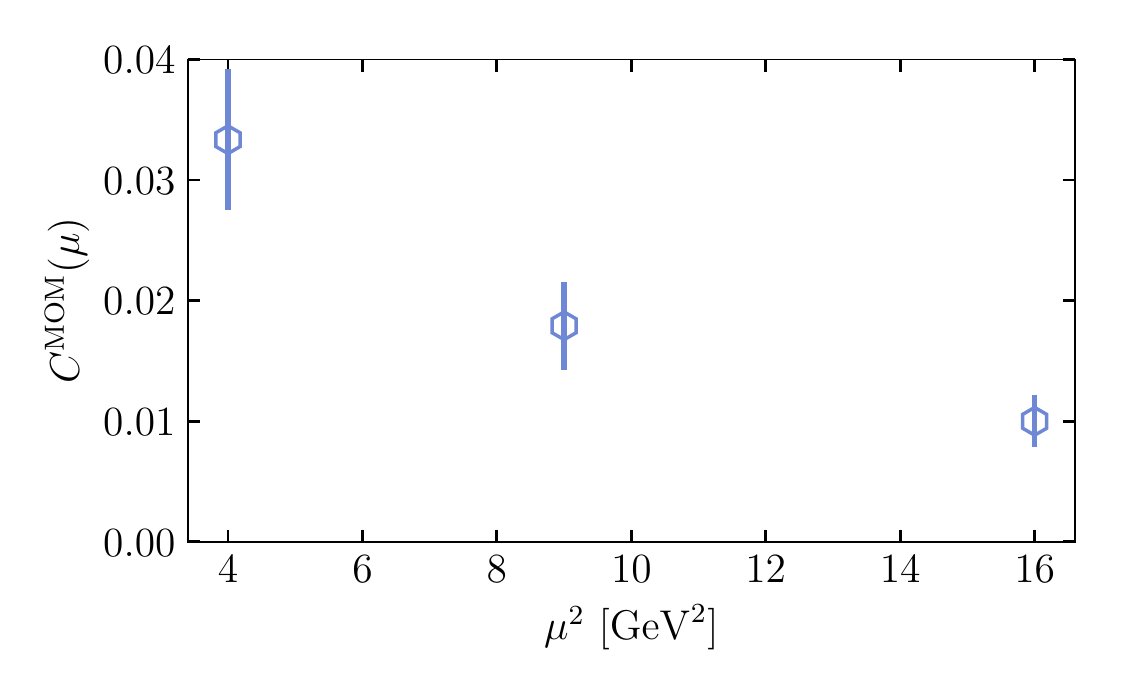}
\caption{The same as Fig.~\ref{fig:correction} but for a correction term, $C^{\text{MOM}}(\mu)$, 
 needed for the tensor current
    renormalisation factor when using the intermediate RI$'$-MOM scheme.}
\label{fig:correctionpm}
\end{figure}

There is more difference between the 2 GeV and the 3 and 4 GeV values in the RI$'$-MOM case
than in the RI-SMOM case. This is reflected in the larger coefficient for the $1/\mu^4$ condensate
term in the fit of -1.19(49). The size of the correction, $C^{\text{MOM}}(\mu)$, needed for $Z_T$ 
when the
RI$'$-MOM scheme is used is shown in Fig.~\ref{fig:correctionpm}. It can be seen
that the correction is larger than for the
RI-SMOM case, because of larger condensate effects. 
It is not surprising that condensate effects are larger in the 
RI$'$-MOM scheme than in RI-SMOM since this has been shown to be true in several other renormalisation 
factors in the past~\cite{Aoki:2007xm,Hatton:2019gha} and is also consistent with the 
mass dependence seen in Fig.~\ref{fig:ZT-mass-dep}. 

\begin{figure}
\centering
\includegraphics[width=0.45\textwidth]{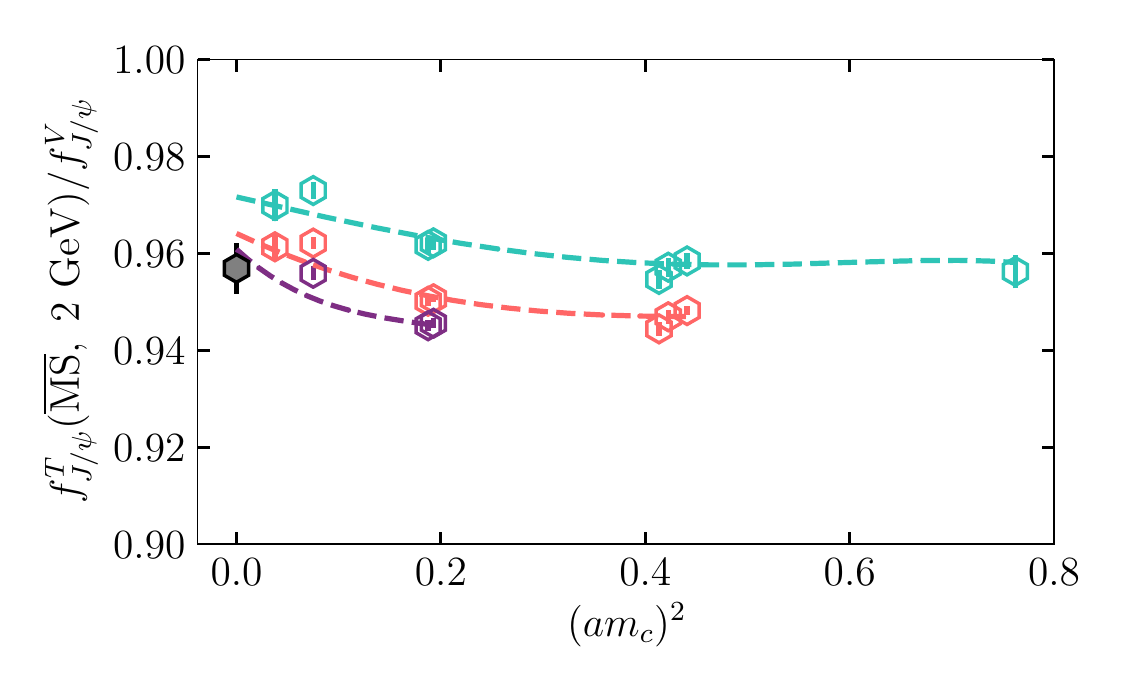}
\caption{Continuum extrapolation of the ratio of the tensor and vector $J/\psi$ decay constants using
    intermediate lattice renormalisation factors in the RI-SMOM scheme.
    Blue points and lines show $\mu=2$ GeV results and fit lines, orange are 3 GeV and purple 4 GeV.
    The bold dashed lines are continuum extrapolations at each $\mu$ value with the condensate and $\alpha_s^4$ terms
    left in. The black hexagon is the continuum extrapolation with condensates and $\alpha_s^4$ errors removed.}
\label{fig:dcratio}
\end{figure}

\begin{figure}
\centering
\includegraphics[width=0.45\textwidth]{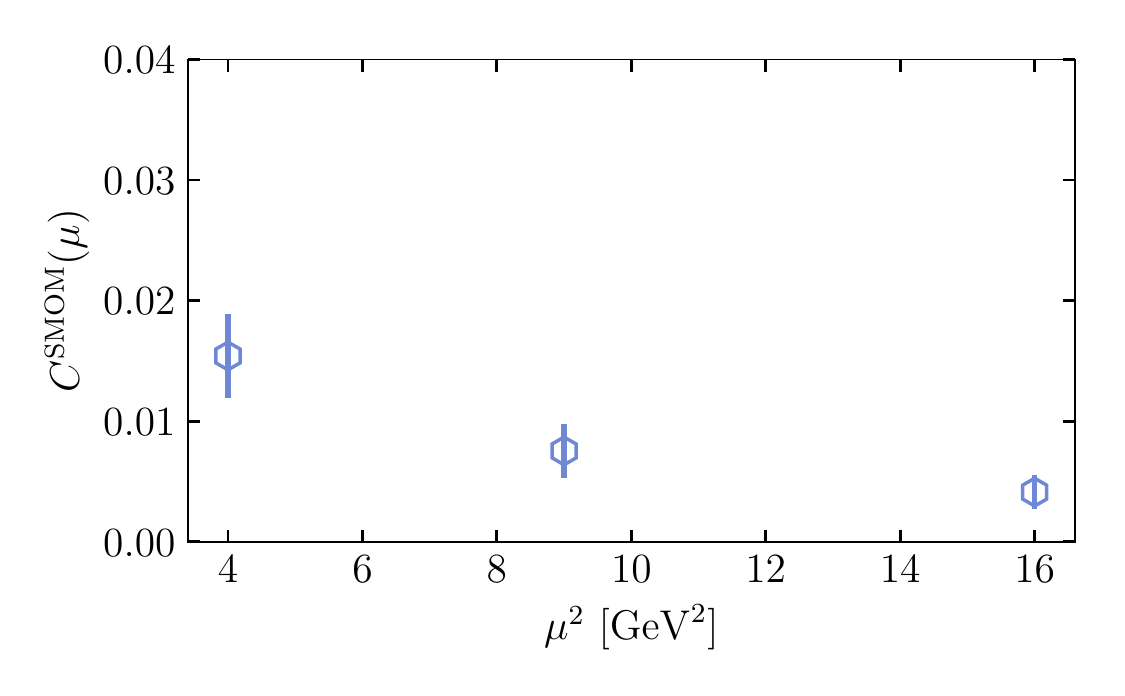}
  \caption{Correction $C^{\mathrm{SMOM}}(\mu)$ for the tensor current renormalisation 
factor, $Z_T$, determined from a fit to the ratio of the $J/\psi$
    tensor and vector decay constants using intermediate 
renormalisation factors in the RI-SMOM scheme.
    This agrees with the results shown in Fig.~\ref{fig:correction}, as expected 
    because the lattice RI-SMOM vector current renormalisation factor has no condensate 
contamination~\cite{Hatton:2019gha}.}
\label{fig:correctionrat}
\end{figure}

\begin{table}
\centering
\caption{Error budget for the ratio of the $J/\psi$ vector and tensor decay
constants. `Statistics', the dominant uncertainty, refers to statistical errors in the amplitudes needed for the decay constants. 
The uncertainties coming from the renormalisation factors, $Z_T$ and $Z_V$, are 
much smaller and are dominated by the contribution from the (doubled) statistical
uncertainties on the low statistics 
ultrafine lattices, set 8. 
The `Missing $\alpha_s^4$' and `Condensates' error contributions
come from the terms in the fit from which the $Z_T$ correction (discussed in the
text) is constructed.}
\label{tab:rat-errbudg}
\begin{tabular}{ll}
\hline \hline
 & $f_{J/\psi}^T/f^V_{J/\psi}$ \\
 \hline
 $(am_c)^2 \to 0$ & 0.11 \\
 $(\tilde{a}\mu)^2 \to 0$ & 0.27 \\
 $Z_T$ & 0.12 \\
 $Z_V$ & 0.14 \\
 Missing $\alpha_s^4$ term & 0.06 \\
 Statistics & 0.41 \\
 Sea mistuning & 0.04 \\
 Condensates & 0.07 \\
 \hline
 Total & 0.54 \\
\hline \hline
\end{tabular}
\end{table}

Since the discretisation effects in $f_{J/\psi}^T$ are similar to those in
$f_{J/\psi}^V$ on the same set of gluon field ensembles we expect
to be able to extract the ratio of the two decay constants 
to a higher precision than can be obtained from
the individual quantities. We may also be able to see a 
clearer indication of the
size of nonperturbative effects in the ratio. 

We show the ratio of $f^T/f^V$ in Fig.~\ref{fig:dcratio} using $Z_T$ 
and $Z_V$ determined in the RI-SMOM scheme. 
We neglect any correlations between the raw values of the decay constants on
each lattice ensemble because the statistical uncertainties 
are so small. We fit 
the values of the ratio to 
Eq.~\eqref{eq:fit-form} and obtain a result for the ratio in 
the continuum limit with
nonperturbative contamination effects removed of:
\begin{equation}
\label{eq:ratvalue}
    \frac{f_{J/\psi}^T (\overline{\text{MS}},2\ \mathrm{GeV})}{f^V_{J/\psi}} = 0.9569(52) \ \ \mathrm{(int.\,SMOM)}.
\end{equation}
The fit has a $\chi^2/\mathrm{dof}$ of 0.2. 

As discussed in
\cite{Hatton:2019gha} the RI-SMOM $Z_V$ contains no 
nonperturbative contamination because of the protection of the Ward-Takahashi
identity and likewise no perturbative matching of SMOM to $\overline{\text{MS}}$ is 
needed. Therefore the condensate and $\alpha_s^4$ terms returned by the fit to
the ratio of the tensor and vector $J/\psi$ decay constants should agree with
those from the fit to just the tensor decay constant. We find that this is the
case for each coefficient individually and for the $Z_T$ correction factor
obtained from their combination which we show for the ratio fit in
Fig.~\ref{fig:correctionrat}.

Because the RI$'$-MOM determination of $Z_V$ has condensate contamination 
(since it is not protected by a Ward-Takahashi identity~\cite{Hatton:2019gha}) 
and perturbative matching is needed to reach $\overline{\text{MS}}$ 
we cannot perform the same analysis for that case. 

We give an error budget for our result for the decay constant ratio
$f_{J/\psi}^T/f^V_{J/\psi}$ in Table~\ref{tab:rat-errbudg}. We can leverage 
this ratio and the vector decay constant determined in~\cite{Hatton:2020qhk} 
to get a slightly more precise value of the tensor decay constant:
\begin{equation}
\label{eq:finvalsmom}
f^T_{J/\psi}(\overline{\text{MS}},2\ \mathrm{GeV}) = 0.3927(27)\ \mathrm{GeV}\ \ (\mathrm{int. \,SMOM}) .
\end{equation}

The vector decay constant result of \cite{Hatton:2019gha} includes 
QED effects from the non-zero electric charge of the valence charm quarks. 
We have not included any electromagnetic effects here. However, the QED
effect on the vector decay constant was at the 0.2\% level and we expect some cancellation 
of these effects in the decay constant ratio, so we neglect these effects here.

\begin{figure}
\centering
\includegraphics[width=0.45\textwidth]{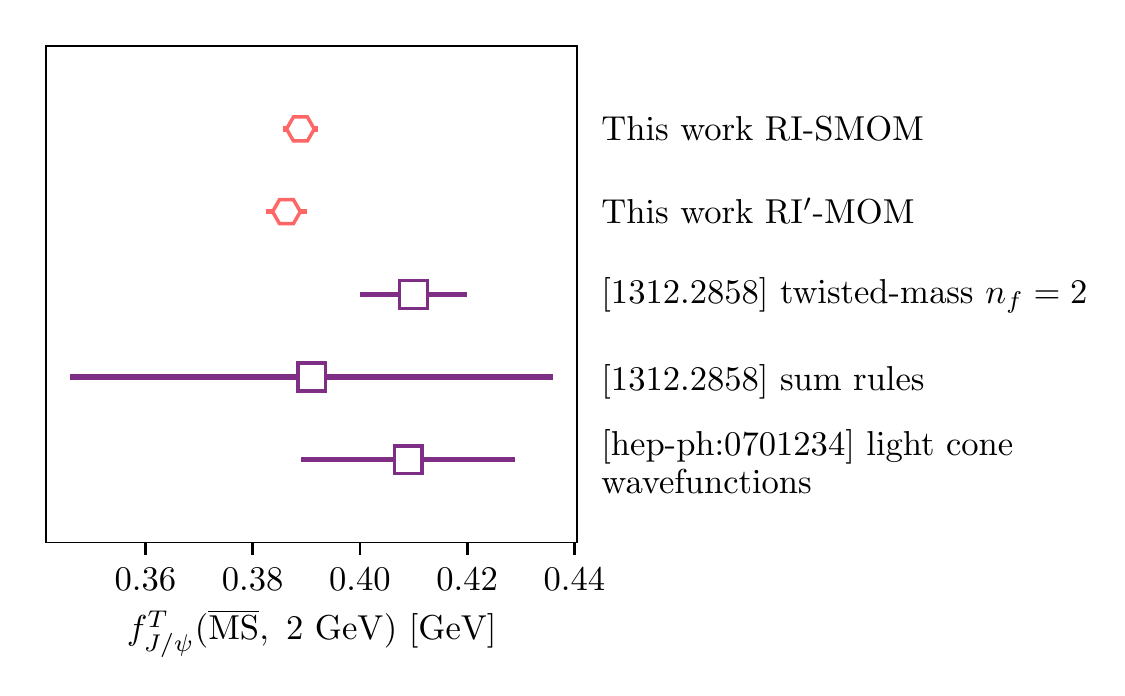}
  \caption{A comparison plot of results for the tensor $J/\psi$ decay constant in the $\overline{\text{MS}}$ 
scheme at a scale of 2 GeV. The top two results are from this work (using both 
  RI-SMOM (Eq.~\eqref{eq:finvalsmom}) and RI$'$-MOM (Eq.~\eqref{eq:finvalpmom}) 
intermediate schemes) and then we include results from~\cite{Becirevic:2013bsa} 
  and~\cite{Braguta:2007fh}.}
\label{fig:fT-comp}
\end{figure}

\begin{figure}
\centering
\includegraphics[width=0.45\textwidth]{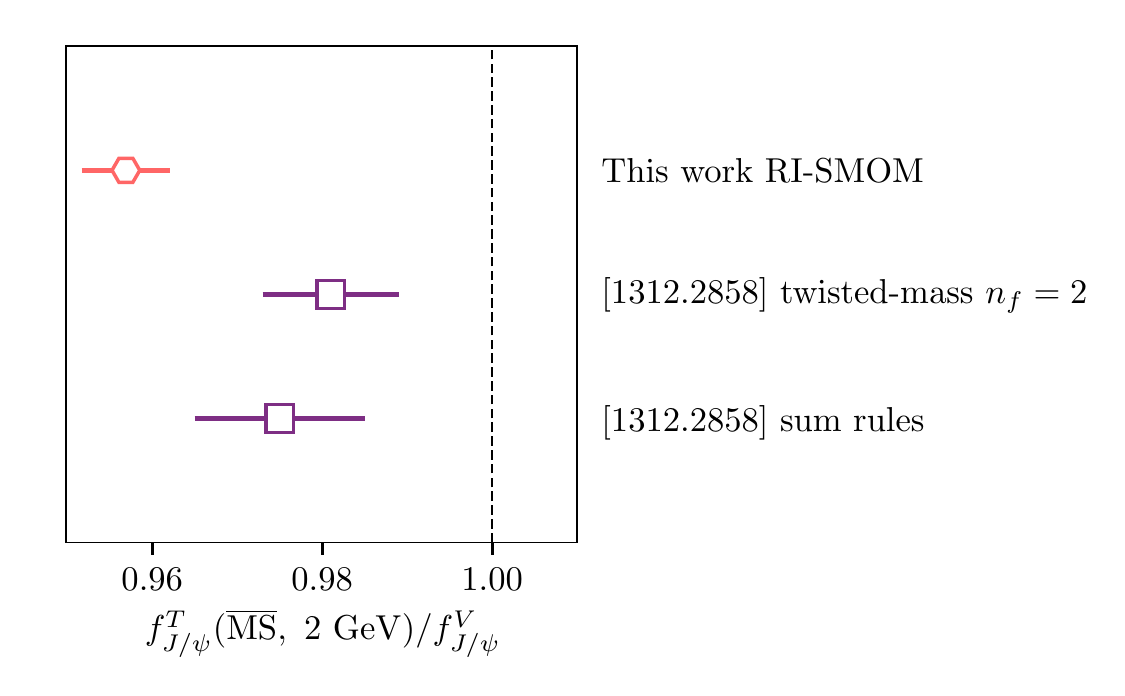}
  \caption{A comparison plot of results for the ratio of tensor and vector 
$J/\psi$ decay constants. The upper result is from this work (Eq.~\eqref{eq:ratvalue}) 
using the RI-SMOM intermediate 
scheme and gives a value significantly below 1 (marked with the black dashed line) 
for this ratio. The lower two results are 
  from~\cite{Becirevic:2013bsa}.}
\label{fig:fratio-comp}
\end{figure}

\section{Discussion: $f_{J/\psi}^T$} \label{sec:discussion-f}

As discussed in Section~\ref{sec:intro} there is no experimental observable available to which 
we can compare 
our tensor current decay constant value. Theoretical results using light-cone 
wavefunctions were presented in~\cite{Braguta:2007fh} and using QCD sum rules 
in~\cite{Becirevic:2013bsa}.
A lattice QCD result using twisted-mass quarks on gluon field ensembles with only $u/d$ quarks in 
the sea ($n_f=2$) was also given in~\cite{Becirevic:2013bsa}. The RI$'$-MOM scheme was 
used to renormalise the lattice tensor and vector currents in that case, 
without studying or removing nonperturbative condensate contamination.
We compare our results to these in Fig.~\ref{fig:fT-comp} where the reduction in uncertainty 
that we have achieved here can clearly be seen.

A comparison plot of values of the decay constant 
ratio $f_{J/\psi}^T(2\, \mathrm{GeV}) / f^V_{J/\psi}$ is shown in Fig.~\ref{fig:fratio-comp}. 
This ratio is expected to be below 1~\cite{Becirevic:2013bsa} but we see that earlier 
results were not able to demonstrate this conclusively. Our value for the ratio is 
8$\sigma$ below 1. 
The value that we obtain for the ratio is just over $1\sigma$ lower 
than the sum rules determination of~\cite{Becirevic:2013bsa} and is over $2\sigma$ lower than the lattice 
QCD result of that work (using their $\sigma$ values). 
In the lattice QCD calculation both the tensor and vector current were renormalised 
in the RI$'$-MOM scheme without accounting for nonperturbative contamination. Our results indicate that this 
could lead to a discrepancy with our results of the size seen. 

\section{Discussion: $Z_T$} \label{sec:discussion}

\begin{table}[b!]
\centering
\caption{$Z_T$ values converting lattice results involving the tensor 
current to the $\overline{\text{MS}}$ scheme, run to a renormalisation scale of
the $b$ quark pole mass. The notation
$Z_T(\mu_1 \vert \mu_2)$ indicates that the intermediate $Z_T^{\text{SMOM}}$ 
has been calculated in the RI-SMOM
scheme at a scale of $\mu_1$ and then converted to the $\overline{\mathrm{MS}}$
scheme and run to a scale of $\mu_2$. 
The superscript denotes that these renormalisation constants have been 
corrected for nonperturbative artefacts and $\alpha_s^4$ errors in $Z_T^{\text{SMOM}}$ 
as described 
in the text. The results with intermediate scales of 2 GeV and 3 GeV then agree well with 
each other and either can be used.  
}
\label{tab:final-ZTs}
\begin{tabular}{lll}
\hline \hline
Set & $Z_T^c(2\ \mathrm{GeV}\ \vert\ m_b)$ & $Z_T^c(3\ \mathrm{GeV}\ \vert\ m_b)$ \\
\hline
vc & 0.9493(42) & - \\
c & 0.9740(43) & 0.9707(25) \\
f & 1.0029(43) & 0.9980(25) \\
sf & 1.0342(43) & 1.0298(25) \\
uf & 1.0476(42) & 1.0456(25) \\
\hline \hline
\end{tabular}
\end{table}

\begin{table*}[t!]
\centering
\caption{Correlation matrix of the corrected $Z_T$ values from Table~\ref{tab:final-ZTs}. These
correlations are large because the matching, running and correction terms are all 
correlated. 
}
\label{tab:ZT-corrs}
\begin{tabular}{llllllllll}
\hline \hline
 & (vc,2) & (c,2) & (f,2) & (sf,2) & (uf,2) & (c,3) & (f,3) & (sf,3) & (uf,3) \\
 \hline
 (vc,2) & 1.0 & 0.99750 & 0.99854 & 0.99475 & 0.93231 & 0.98398 & 0.98611 & 0.98713 & 0.96383 \\
 (c,2) & 0.99750 & 1.0 & 0.99777 & 0.99430 & 0.93294 & 0.98314 & 0.98371 & 0.98487 & 0.96243 \\
 (f,2) & 0.99854 & 0.99777 & 1.0 & 0.99605 & 0.93562 & 0.98045 & 0.98323 & 0.98423 & 0.96263 \\
 (sf,2) & 0.99475 & 0.99430 & 0.99605 & 1.0 & 0.93197 & 0.97361 & 0.97632 & 0.98097 & 0.95627 \\
 (uf,2) & 0.93231 & 0.93294 & 0.93562 & 0.93197 & 1.0 & 0.90439 & 0.90777 & 0.90941 & 0.96855 \\
 (c,3) & 0.98398 & 0.98314 & 0.98045 & 0.97361 & 0.90439 & 1.0 & 0.99909 & 0.99807 & 0.96824 \\
 (f,3) & 0.98611 & 0.98371 & 0.98323 & 0.97632 & 0.90777 & 0.99909 & 1.0 & 0.99868 & 0.96951 \\
 (sf,3) & 0.98713 & 0.98487 & 0.98423 & 0.98097 & 0.90941 & 0.99807 & 0.99868 & 1.0 & 0.96909 \\
 (uf,3) & 0.96383 & 0.96243 & 0.96263 & 0.95627 & 0.96855 & 0.96824 & 0.96951 & 0.96909 & 1.0 \\
 \hline \hline
 \end{tabular}
 \end{table*}

In the discussion presented above in Section~\ref{sec:jpsi-tensor} 
we ran all of our results, after
converting to $\overline{\mathrm{MS}}$, to a common scale of 2 GeV and then
determined and subtracted a correction that depends on $\mu$. 
This correction needs to be applied to our $Z_T$ values for future use. 
The scale of 2 GeV allows us to
compare directly to the results of \cite{Becirevic:2013bsa} in 
Section~\ref{sec:discussion-f}. However, another
scale is useful when computing form factors for semileptonic $B$ decay processes. 
Then differential rates are calculated as functions
of products of the form factors and appropriate Wilson coefficients of the weak
Hamiltonian. These Wilson coefficients are scale dependent and are typically
calculated at a scale equal to the $b$ pole mass, 4.8 GeV, see 
for example \cite{Altmannshofer:2008dz}. 
We therefore present our $Z_T$ values run to this scale. If the $b$ quark 
running mass of 4.2 GeV were used instead of the pole mass then 
the values would be approximately 
1\% larger.

In Table~\ref{tab:final-ZTs} we give the corrected results for
$Z_T$ in the $\overline{\mathrm{MS}}$ scheme at a scale equal to the $b$ quark
pole mass calculated from intermediate values of $Z_T$ in
the RI-SMOM scheme at 2 and 3 GeV. We use a notation
$Z_T(\mu_{\mathrm{SMOM}}\ \vert\ \mu_{\overline{\mathrm{MS}}})$ where
$\mu_{\mathrm{SMOM}}$ is the scale at which the RI-SMOM calculation was
performed and $\mu_{\overline{\mathrm{MS}}}$ is the final scale at which the
$\overline{\mathrm{MS}}$ result is presented. It can be seen that the addition of 
the correction results in $Z_T$ values that agree for different intermediate 
scales once run to the same final scale (this would not be true 
for uncorrected values). We also give the correlations
between these numbers in Table~\ref{tab:ZT-corrs}.

\section{Conclusions} \label{sec:conclusions}

We have shown here that it is possible to renormalise lattice tensor currents 
to give accurate results for continuum matrix elements in the $\overline{\text{MS}}$ 
scheme using nonperturbative determination of intermediate renormalisation factors 
in momentum-subtraction schemes. A key requirement is that the nonperturbative 
renormalisation factors should be obtained at multiple values 
of the renormalisation scale, $\mu$, 
so that $\mu$-dependent nonperturbative (condensate) 
contamination of $Z_T$ can be fitted and removed. This contamination would otherwise 
give a systematic error of 1.5\% using the RI-SMOM scheme and 3\% using the 
RI$'$-MOM scheme in our calculation.  

In order to do this we have determined the $J/\psi$ tensor decay constant, 
$f_{J/\psi}^T$ so that we can study the continuum limit of a tensor current matrix element. 
Using $n_f=2+1+1$ HISQ lattices and the local tensor current, we obtain a 
0.7\%-accurate value for $f_{J/\psi}^T$ of (repeating Eq.~\eqref{eq:finvalsmom}) 
\begin{equation}
\label{eq:finvalsmom2}
f^T_{J/\psi}(\overline{\text{MS}},2\ \mathrm{GeV}) = 0.3927(27)\ \mathrm{GeV}\ \ (\mathrm{int. \,SMOM}) .
\end{equation}
This uses our preferred intermediate RI-SMOM scheme and makes use of the 
determination of the ratio of 
tensor to vector decay constants and the fact that the vector current 
renormalisation is protected 
by the Ward-Takahashi identity in this scheme~\cite{Hatton:2019gha}.  
We also obtain a 0.5\%-accurate value for the ratio itself (repeating Eq.~\eqref{eq:ratvalue}), 
\begin{equation}
\label{eq:ratvalue2}
    \frac{f_{J/\psi}^T (\overline{\text{MS}},2\ \mathrm{GeV})}{f^V_{J/\psi}} = 0.9569(52) \ \ (\mathrm{int.\,SMOM}) .
\end{equation}
This shows unequivocally that the ratio is less than 1. 

Finally, in Tables~\ref{tab:final-ZTs} and~\ref{tab:ZT-corrs}, we give $Z_T$ renormalisation 
factors that can be used, for example, in a future 
determination (underway) of the tensor form factor for the 
rare flavour-changing neutral current process $B\rightarrow K \ell^+\ell^-$ using HISQ quarks. 
These $Z_T$ values take results determined with the local HISQ lattice tensor 
current and convert them into values in the $\overline{\text{MS}}$ scheme at the 
scale of $m_b$, to be multiplied by Wilson coefficients from the effective weak Hamiltonian 
determined at this scale. We have corrected these $Z_T$ values so that 
they are free of the systematic error from condensate contamination 
of the intermediate momentum-subtraction scheme.

\subsection*{\bf{Acknowledgements}}

We are grateful to the MILC collaboration for the use of
their configurations and code. 
Computing was done on the Cambridge service for Data 
Driven Discovery (CSD3), part of which is operated by the 
University of Cambridge Research Computing on behalf of 
the DIRAC 
HPC Facility of the Science and Technology Facilities 
Council (STFC). The DIRAC component of CSD3 was funded by 
BEIS capital funding via STFC capital grants ST/P002307/1 and 
ST/R002452/1 and STFC operations grant ST/R00689X/1. 
DiRAC is part of the national e-infrastructure.  
We are grateful to the CSD3 support staff for assistance.
Funding for this work came from the UK
Science and Technology Facilities Council grants 
ST/L000466/1 and ST/P000746/1 and from the National Science 
Foundation.

\bibliography{ZTpaper}

\end{document}